\documentclass[twocolumn]{aastex631}
\usepackage{isotope}
\usepackage{units}
\usepackage{hyperref}
\usepackage{amsmath}
\usepackage{soul}
\submitjournal{ApJ}

\shorttitle{Light neutron-capture elements in neutrino-driven outflows}
\shortauthors{Psaltis et al.}
\graphicspath{{./}{figures/}}

\begin{document}

\title{Neutrino-Driven Outflows and the Elemental Abundance Patterns of Very Metal-Poor Stars}

\author[0000-0003-2197-0797]{A. Psaltis}
\affiliation{Department of Physics, North Carolina State University,
Raleigh, NC, 27695, USA}
\affiliation{Triangle Universities Nuclear Laboratory, Duke University, Durham, NC, 27710, USA}
\email{psaltis.tha@duke.edu}

\author[0000-0001-8168-4579]{M. Jacobi}
\affiliation{Institut f\"ur Kernphysik, Technische Universit\"at Darmstadt,
Schlossgartenstr. 2, Darmstadt 64289, Germany}

\author[0000-0001-9849-5555]{F. Montes}
\affiliation{Facility for Rare Isotope Beams, East Lansing, MI, 48824, USA}
\affiliation{Joint Institute for Nuclear Astrophysics – CEE, Michigan State
University, East Lansing, Michigan 48824, USA}

\author[0000-0002-6995-3032]{A. Arcones}
\affiliation{Institut f\"ur Kernphysik, Technische Universit\"at Darmstadt,
Schlossgartenstr. 2, Darmstadt 64289, Germany}
\affiliation{GSI Helmholtzzentrum f\"ur Schwerionenforschung GmbH, Planckstr. 1, Darmstadt 64291, Germany}
\email{almudena.arcones@physik.tu-darmstadt.de}

\author[0000-0002-7277-7922]{C. J. Hansen}
\affiliation{Institute for Applied Physics, Goethe University Frankfurt, Max-von-Laue-Str. 12, Frankfurt am Main 60438, Germany}

\author[0000-0003-1674-4859]{H. Schatz}
\affiliation{Facility for Rare Isotope Beams, East Lansing, MI, 48824, USA}
\affiliation{Joint Institute for Nuclear Astrophysics – CEE, Michigan State
University, East Lansing, Michigan 48824, USA}
\affiliation{Department of Physics and Astronomy, Michigan State University, 567 Wilson Road, East Lansing, MI 48824, USA}



\begin{abstract}
The elemental abundances between strontium and silver ($Z = 38-47$) observed in the atmospheres of very metal-poor stars (VMP) in the Galaxy may contain the fingerprint of the weak $r$-process and $\nu p$-process occurring in early core-collapse supernovae explosions. In this work, we combine various astrophysical conditions based on a steady-state model to cover the richness of the supernova ejecta in terms of entropy, expansion timescale, and electron fraction. The calculated abundances based on different combinations of conditions are compared with stellar observations with the aim of constraining supernova ejecta conditions. We find that some conditions of the neutrino-driven outflows consistently reproduce the observed abundances of our sample. In addition, from the successful combinations, the neutron-rich trajectories better reproduce the observed abundances of Sr-Zr ($Z= 38-40$), while the proton-rich ones, Mo-Pd ($Z= 42-47$).
\end{abstract}



\section{Introduction} \label{sec:intro}

The origin of elements heavier than iron (Z= 26) in the cosmos is one of the most intriguing open questions in nuclear astrophysics. Roughly half of them have traditionally been attributed to the rapid-neutron capture process~\citep[\textit{r}-process;][]{horowitz2019r, cowan2021origin} however, its astrophysical site or sites are still disputed, and it is unclear whether other processes such as the \textit{$\nu$p}- or the \textit{i}-process contribute as well. A recently confirmed \textit{r}-process site is the merging of two neutron stars (NSMs), which was detected both via gravitational waves ~\citep[GW170817;][]{abbott2017gw170817}, and its electromagnetic follow-up transient -- kilonova ~\citep[AT2017gfo][]{drout2017light}.
Despite these successes, due to their rather rare nature and long delay times, NSMs appear to be unable to fully account for the evolution of \textit{r}-process abundances in the Galaxy~\citep[see e.g.,][]{cote2019neutron, 2021MNRAS.503....1C, 2023ApJ...943L..12K}. Other astrophysical sites, such as magnetorotational supernovae ~\citep{Winteler:2012, Nishimura2017, reichert2021nucleosynthesis, 2023MNRAS.518.1557R}, may also produce heavy elements up to the third \textit{r}-process peak and the actinides. Detection of $\isotope[244][]{Pu}$ and $\isotope[60][]{Fe}$ in deep-sea crust sediments on Earth suggests a core-collapse supernovae contribution for these specific isotopes. However, NSMs can also account for the production of $\isotope[244][]{Pu}$~\citep{2021Sci...372..742W, 2023ARNPS..73..365F}.

Moreover, neutrino-driven supernovae may explain observations that indicate an early enrichment of the interstellar medium with elements up to around silver, just before the second \textit{r}-process peak~\cite{arcones2011production}. In this paper, we use these lighter heavy elements between strontium and silver (Z= 38 -- 47) to constrain supernova conditions by comparing calculated and observed abundances. Spectroscopic studies of very metal-poor stars show a general robustness not only with their \textit{r}-process abundances but also with the solar \textit{r}-process pattern\footnote{Solar \textit{r}-process abundances are calculated by subtracting the \textit{s}- and \textit{p}-process contributions from the total abundance pattern~\citep{1999A&A...342..881G, sneden2008neutron}. Recently~\cite{2020MNRAS.491.1832P} proposed a novel approach by using Galactic Chemical Evolution (GCE) models.}, especially in the lanthanide region~\citep[see for example;][]{2000ApJ...533L.139S, Hill2002, 2007A&A...476..935F, 2014AJ....147..136R, 2018ApJ...854L..20S, 2020ApJ...898...40C, 2022ApJS..260...27R}. This robustness does not extend to the lighter heavy elements,  between strontium and silver (Z = 38-47), where there is a consistent scatter~\citep[see for example Figures 11 and 7 from][respectively]{sneden2008neutron, 2010A&A...516A..46M}. Moreover, there are some stars with high abundances of lighter heavy elements relative to heavy r-process elements between the second and third \textit{r}-process peaks. These stars have a high ratio of \textit{r}-process Sr/Eu (see in Table~\ref{tab:2} the so called Honda star, HD~122563) compared to the solar \textit{r}-process -- $\mathrm{\log \epsilon(Sr/Eu)}$= 1.46~\citep{sneden2008neutron}. Here, we assume that this enhancement may be due to a nucleosynthesis contribution different from the main \textit{r}-process which is responsible for abundances up to the third \textit{r}-process peak. This additional contribution has been thoroughly discussed in the literature~\citep[][for some notable examples]{1996ApJ...466L.109W, 2000PhR...333...77Q, 2002PASP..114.1293T, 2005ApJ...632..611A, 2006ApJ...641L.117O, montes2007, 2007PhR...442..237Q, hansen2014many, 2014ApJ...787...10B, 2015ApJ...801...53C} and operates very early in Galactic evolution. Possible candidates are the weak \textit{r}-process~\citep[also known as $\alpha$-process;][]{woosley1992alpha, witti1994nucleosynthesis, qian1996nucleosynthesis, hoffman1997nucleosynthesis, wanajo2001r,  arcones2011production, 2013ApJ...770L..22W, hansen2014many, bliss2017impact, bliss2018survey}, and the $\nu p$-process~\citep{frohlich2006neutrino, wanajo2006rp, pruet2006nucleosynthesis, 2011ApJ...729...46W, nishimura2019uncertainties} both occurring in neutrino-driven outflows of explosive astrophysical environments, such as core-collapse supernovae.

In this work, we survey the astrophysical conditions of the neutrino-driven outflows in supernovae ejecta, both neutron-rich and proton-rich, using an extensive library of thermodynamic trajectories that span the relevant parameter space of astrophysical conditions. We use superpositions of different conditions and compare the resulting nucleosynthesis with observations of very metal-poor stars that show an overproduction of first $r$-process peak isotopes compared to Eu. We aim to pinpoint astrophysical conditions that are favorable to replicating the observed abundances of heavy-element abundances between Sr and Ag in very metal-poor stars.

This paper is structured as follows: in \S~\ref{sec:astro} we present the astrophysical conditions we used in the present study. In \S~\ref{sec:stars} we discuss the very metal-poor star sample we use to compare our nucleosynthesis calculations to. In \S~\ref{sec:methods} we present our procedure to compare observations with combinations of astrophysical conditions and in \S~\ref{sec:results} we show our results. Finally, in \S~\ref{sec:conclusions} we present our conclusions and a discussion.

\section{Astrophysical conditions}
\label{sec:astro}

In this section, we discuss two different nucleosynthesis processes that occur in neutrino-driven outflows, namely the weak \textit{r}- and the $\nu p$-process~\citep[see][and references therein]{2013JPhG...40a3201A}.

Both nucleosynthesis processes start from very hot material ($T > 10$~GK) ejected from the surface of the nascent proto-neutron star (PNS) by neutrinos. This material is composed of dissociated nucleons~\citep[see Figure 1 of][for a schematic]{PhysRevC.106.045805}. Depending on the electron neutrino and electron anti-neutrino energies and luminosities and on how fast the matter expands, the ejected material can be neutron- or proton-rich. Core-collapse supernovae are complex and asymmetric and conditions evolve rapidly with time.  Therefore, their ejecta are a mixture of materials that encounter different types of conditions. Current simulations show that neutrinos drive most of the matter to proton-rich conditions. However, some bubbles or pockets of material expand fast and stay slightly neutron-rich~\citep[for example ][]{Eichler2018, Harris2017, Wanajo2018, OConnor:2015, muller2017, bollig2021self, 2021Natur.589...29B, Navo2023}.

\begin{figure}[hbpt!]
    \centering
    \includegraphics[width=0.5\textwidth]{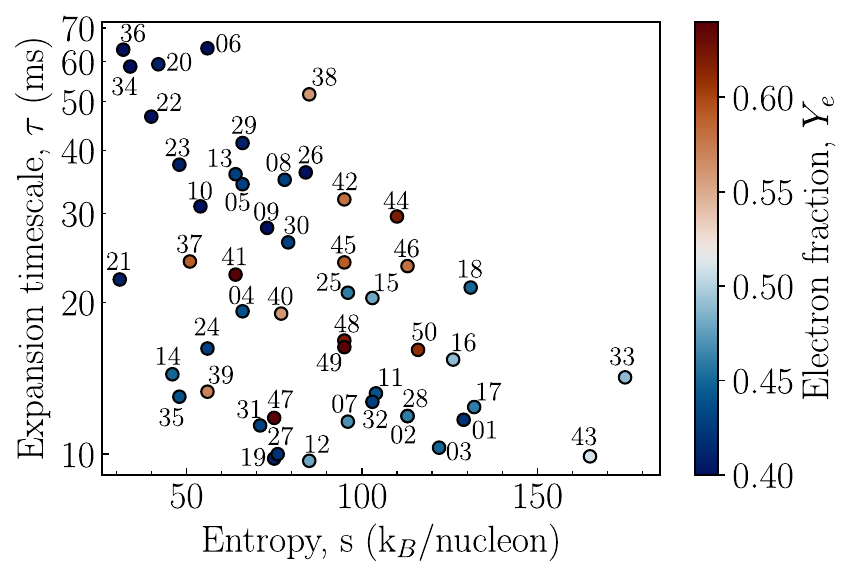}
    \caption{Astrophysical conditions in the entropy per baryon $s$ and expansion timescale $\tau$ space. The color-code corresponds to the electron fraction $Y_e$.}
\label{fig:1}
\end{figure}

The nucleosynthesis of neutrino-driven outflows can be determined from thermodynamic trajectories, characterized by their expansion timescale ($\tau$), entropy ($s$) and electron fraction ($Y_e$)~\cite[see][]{qian1996nucleosynthesis, arcones2014}. Each trajectory provides the time evolution of temperature and density, as well as neutrino energies and luminosities, which are used to calculate the associated nucleosynthesis with a nuclear reaction network. According to the electron fraction $Y_e$, nucleosynthesis can be  the weak \textit{r}- ($Y_e < 0.5$) or the $\nu p$-process ($Y_e > 0.5$). In previous works, we focused on individual conditions analyzing single trajectories \citep{bliss2018survey, arcones2011production, 2022ApJ...935...27P}. To account for the diversity of conditions in supernovae, in this work we explore a total of 50, either neutron-rich or proton-rich conditions to explain the $Z= 38-47$ abundance pattern of Galactic metal-poor stars. In Figure~\ref{fig:1}, we present those trajectories in the entropy per baryon-expansion timescale-$Y_e$ phase-space at around 10~GK (when we start the associated nucleosynthesis using a nuclear reaction network): 36 trajectories from~\cite{bliss2020nuclear} are neutron-rich and the rest are proton-rich.

\subsection{Weak \textit{r}-process}
\label{sec:weak_r}

In moderate neutron-rich outflows, as the ejecta expand and cool down, mainly iron-peak nuclei are produced via the nuclear statistical equilibrium (NSE). The main reaction channel that links the nucleosynthesis from $A<5$ nuclei across the $A=5$ and $A=8$ stability gaps to the CNO element region and beyond is $\alpha + \alpha + n \rightarrow \isotope[9][]{Be}$, followed by the $\isotope[9][]{Be}(\alpha,n)\isotope[12][]{C}$. At $T \approx 5$~GK, NSE cannot be maintained anymore and an $\alpha$-rich freeze-out occurs. Nucleosynthesis proceeds mainly through charged-particle reactions -- $(\alpha, n), (p,n), (\alpha, \gamma)$ -- and neutron captures $(n,\gamma)$. When the temperature falls to $\approx1.5-3$~GK  the $\alpha$-induced reactions become too slow and the nucleosythesis of heavier nuclei ceases. This scenario is known as the weak \textit{r}-process or $\alpha$-process~\citep{woosley1992alpha, hoffman1997nucleosynthesis, arcones2011production}.

\citet{bliss2018survey} explored the relevant parameter space for conditions in the weak $r$-process ($Y_e$, entropy per baryon, and the expansion timescale) using a steady-state model following~\cite{2000ApJ...533..424O}. Subsequent sensitivity studies by~\citet{bliss2020nuclear} and~\citet{2022ApJ...935...27P} tested the importance of $(\alpha, n)$ reaction rates in the final abundance patterns based on 36 representative trajectories from the CPR2 group\footnote{The representative trajectories can be found in~\citep{nuc-astro}.} -- the conditions that produce elements with $Z = 38-47$. These reactions were found to be the main nuclear physics uncertainty in weak \textit{r}-process nucleosynthesis calculations ~\citep{bliss2017impact,bliss2020nuclear}.~\cite{2022ApJ...935...27P} went a step further and explored the impact of nuclear physics uncertainties on predictions of elemental abundance ratios and compared them with observations of Galactic metal-poor stars. Both studies have motivated several experimental studies in the nuclear astrophysics community~\citep[for example]{2021ApJ...908..202K, 2021PhRvC.104c5804S, 2023EPJWC.27911003A}.

\subsection{$\nu p$-process}
\label{sec:nup}

The $\nu p$-process operates in proton-rich outflows  ($Y_e > 0.5$) and can create lighter heavy elements up to Ag \citep{Froehlich:2006b,wanajo2006rp,pruet2006nucleosynthesis}.
Furthermore, the $\nu p$-process can be a mechanism for the production of the lighter \textit{p}-nuclei~\citep{2011ApJ...729...46W, 2013RPPh...76f6201R, arcones2011production}.

The most abundant isotope during the $\nu p$-process is \isotope[56]{Ni} which is already reached during NSE by a series of $\alpha$ and $p$ captures starting with the triple $\alpha$ reaction, $\alpha(2\alpha, \gamma)\isotope[12]{C}$.
After the matter drops out of NSE, the abundance distribution is mainly determined by a $(p,\gamma)$-$(\gamma, p)$ equilibrium, due to the large abundance of free protons. Matter gets accumulated in bottlenecks, mainly \isotope[56]{Ni} and \isotope[64]{Ge}, due to their long $\beta$-decay lifetimes. However, electron-antineutrino captures on free protons produce enough neutrons to overcome the bottlenecks by (n,p) reactions \citep{Froehlich:2006b, wanajo2006rp, pruet2006nucleosynthesis}. Therefore, the production of elements beyond iron by the $\nu p$-process depends on the  flux of antineutrinos and  the abundance of free protons. The efficiency of the process is given by the ratio between free neutrons generated by electron antineutrino absorption on protons and seed nuclei~\citep{pruet2006nucleosynthesis}:
\begin{equation}\label{eq:Delta_n}
\Delta_n = \frac{Y_p}{Y_{\text{seed}}} \int_{T_9 < 3} \lambda_{\bar{\nu}_e} \,\mathrm{d}t\,,
\end{equation}
where, $Y_p$ and $Y_{\text{seed}}$ represent the proton and seed nuclei abundances, respectively, and $\lambda_{\bar{\nu}_e}$ denotes the rate of antineutrino captures on free protons ($\lambda_{\bar{\nu}} \propto L_{\bar{\nu}} \langle \varepsilon_{\bar{\nu}}\rangle / r^2$, where $L_{\bar{\nu}}$ is the electron-antineutrino energy luminosity, $\langle\varepsilon_{\bar{\nu}_e}\rangle$ their average energy, and $r$ the radius). The efficiency of the $\nu p$-process  and thus $\Delta_n$ increases for larger $Y_p/Y_{\text{seed}}$  and higher antineutrino energies and luminosities. The proton-to-seed ratio is larger for higher entropies and electron fractions. Typical values of $\Delta_n$ for ejecta with $Y_e \lesssim 0.65$  range from 1 to 100.

The astrophysical and nuclear physics uncertainties of the $\nu p$-process have been explored extensively by~\cite{2011ApJ...729...46W, 2012ApJ...750...18A, nishimura2019uncertainties} and motivated new measurements of nuclear masses and thermonuclear reaction rates~\citep[e.g., ][]{2011PhRvC..84d5807F, 2018PhLB..781..358X, 2021PhRvC.104d2801R, PhysRevLett.129.162701}.

Similar to \citet{bliss2018survey}, we calculate neutrino-driven wind trajectories for proton-rich conditions. We vary the electron fraction at $T=\unit[10]{GK}$ between 0.51 and 0.65 and  assume that the number luminosities are the same for electron neutrino and antineutrino.  We use constant antineutrino energy luminosity $L_{\bar{\nu}_e} = \unit[3 \times 10^{51}]{erg s^{-1}}$ and antineutrino energy $\langle\varepsilon_{\bar{\nu}_e}\rangle = \unit[16.66]{MeV}$. Note, that increasing the average energy or number luminosity of antineutrinos while keeping the electron fraction constant increases $\Delta_n$ and thus leads to the formation of heavier nuclei. However, the effect is almost equivalent to changing $\Delta_n$ by other means (\textit{e.g.}, by increasing the entropy or electron fraction). It is thus sufficient for our purpose to only consider one value for the antineutrino luminosity and average energy.

\begin{figure}[hbpt!]
    \centering
    \includegraphics[width=.5\textwidth]{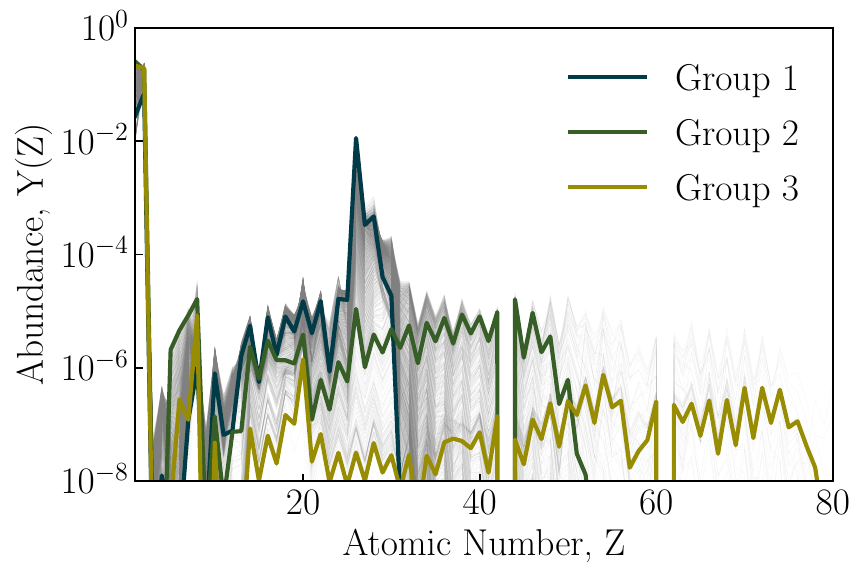}
    \includegraphics[width=.5\textwidth]{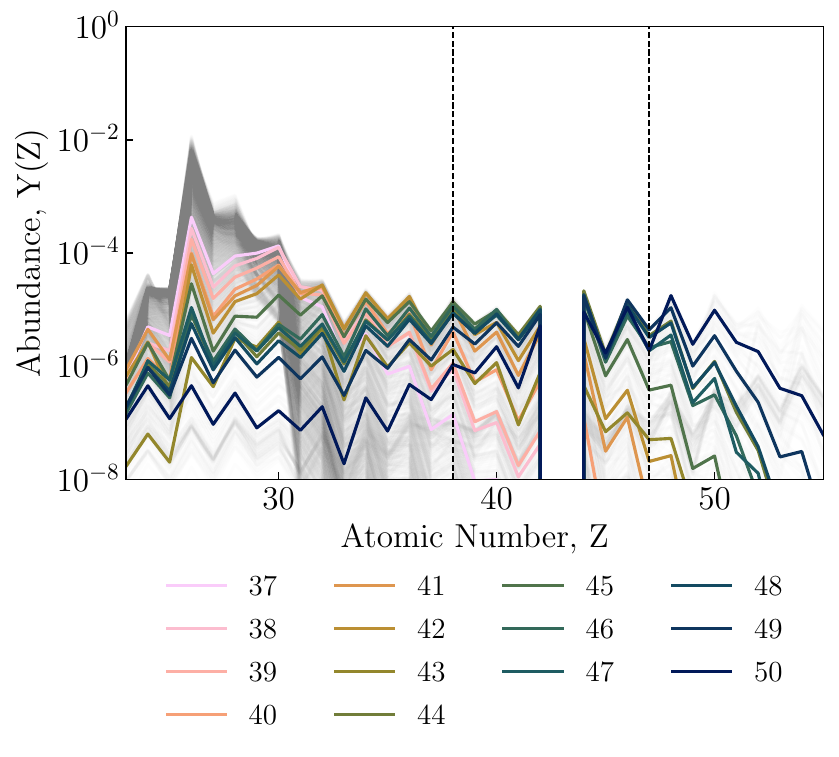}
    \caption{(Top) Abundance patterns for the proton-rich trajectories ($Y_e>0.5)$. A representative pattern for each of the three groups mentioned in the text is shown in color. (Bottom) Abundance patterns for the 14 trajectories used in the present study. The strontium-silver (Z= 38-47) region is indicated by the dashed vertical lines.}
    \label{fig:2}
\end{figure}

Figure~\ref{fig:2} shows the abundance patterns of all the $Y_e>0.5$ trajectories. These patterns can be sorted into three groups: mainly iron-peak nuclei without any significant production of $Z>30$ elements, patterns showing production of elements between strontium and silver, and patterns that can produce elements up to around $Z=80$. The second group, from which we selected fourteen representative trajectories (37-50 in Table~\ref{tab:1}, and shown in the lower panel of Figure~\ref{fig:2}), produces elements between strontium and silver. The trajectories responsible for the third group (with abundances up to around $Z=80$) are too extreme and are probably rarely or not found in ccSNe explosions. The proton-rich trajectories have a smaller variability in the final abundance pattern than the neutron-rich ones. The reaction flow of the $\nu p$-process is much more constrained compared to the weak \textit{r}-process, which operates on the neutron-rich side of the stability valley~\citep{arcones2011production}.
For this reason, in Table~\ref{tab:1} we select only 14 proton-rich conditions, compared to 36 neutron-rich conditions.

\startlongtable
\begin{deluxetable}{ccccc}
\tablecaption{Main Astrophysical Conditions for the Trajectories used in the Present Study. \label{tab:1}}
\tablehead{\colhead{Trajectory} & \colhead{$Y_e$} & \colhead{$s$ ($k_B$/nucleon)} &  \colhead{$\tau$ (ms)} & \colhead{$\Delta_n$}}
\startdata
01 & 0.42 & 129 & 11.7  & \nodata\\
02 & 0.45 & 113 & 11.9  & \nodata\\
03 & 0.45 & 122 & 10.3 & \nodata\\
04 & 0.44 & 66 & 19.2 & \nodata \\
05 & 0.43 & 66 & 34.3 & \nodata \\
06 & 0.40 & 56 & 63.8 & \nodata  \\
07 & 0.47 & 96 & 11.6 & \nodata  \\
08 & 0.43 & 78 & 35.0 & \nodata  \\
09 & 0.40 & 73 & 28.1 & \nodata \\
10 & 0.40 & 54 & 31.0 & \nodata  \\
11 & 0.44 & 104 & 13.2 & \nodata\\
12 & 0.48 & 85 & 9.7 & \nodata   \\
13 & 0.43 & 64 & 35.9 & \nodata  \\
14 & 0.45 & 46 & 14.4  & \nodata\\
15 & 0.48 & 103 & 20.4  & \nodata\\
16 & 0.49 & 126 & 15.4  & \nodata\\
17 & 0.46 & 132 & 12.4  & \nodata\\
18 & 0.45 & 131 & 21.4  & \nodata\\
19 & 0.41 & 75 & 9.8  & \nodata\\
20 & 0.41 & 42 & 59.3 & \nodata \\
21 & 0.41 & 31 &  22.2 & \nodata\\
22 & 0.40 & 40 & 46.7 & \nodata  \\
23 & 0.41 & 48 & 37.5 & \nodata \\
24 & 0.43 & 56 & 16.2 & \nodata \\
25 & 0.46 & 96 & 20.9 & \nodata  \\
26 & 0.40 & 84 & 36.2 & \nodata  \\
27 & 0.42 & 76 & 10.0 & \nodata \\
28 & 0.46 & 113& 11.9 & \nodata  \\
29 & 0.41 & 66 & 41.4 & \nodata  \\
30 & 0.43 & 79 & 26.3 & \nodata  \\
31 & 0.43 & 71 & 11.4 & \nodata  \\
32 & 0.43 & 103 & 12.7  & \nodata\\
33 & 0.49 & 175 & 14.2 & \nodata \\
34 & 0.40 & 34 & 58.7  & \nodata\\
35 & 0.44 & 48 & 13.0  & \nodata\\
36 & 0.40 & 32 & 63.4  & \nodata\\
37 & 0.59 & 51 & 24.1 & 3.4 \\
38 & 0.56 & 85 & 51.7 & 5.5 \\
39 & 0.57 & 56 & 13.3 & 5.9 \\
40 & 0.56 & 77 & 19.0 & 8.8 \\
41 & 0.64 & 64 & 22.7 & 11.0 \\
42 & 0.58 & 95 & 32.0 & 14.8 \\
43 & 0.51 & 165 & 9.9 & 16.8 \\
44 & 0.62 & 110 & 29.6 & 35.3 \\
45 & 0.59 & 95 & 24.0 & 21.6 \\
46 & 0.58 & 113 & 23.6 & 31.1 \\
47 & 0.64 & 75 & 11.8 & 32.1 \\
48 & 0.62 & 95 & 16.8 & 34.6 \\
49 & 0.64 & 95 & 16.3 & 44.2 \\
50 & 0.61 & 116 & 16.1 & 57.4 \\
\enddata
\tablecomments{$\Delta_n$ is defined only for trajectories with $Y_e>0.5$ ($\nu p$-process).}
\end{deluxetable}

\section{Abundance observations from metal--poor stars}
\label{sec:stars}

\begin{figure}[hbpt!]
    \centering
    \includegraphics[width=0.5\textwidth]{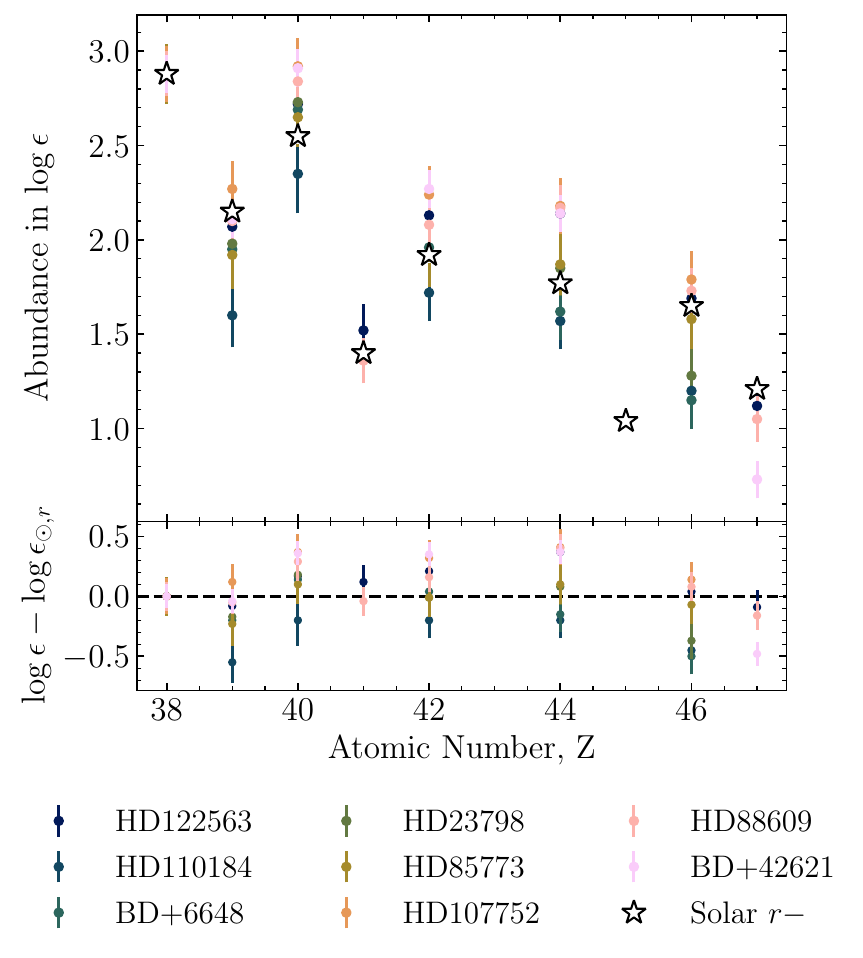}
    \caption{Elemental abundances of the stellar sample used in this study \textit{normalized }to the solar $r-$process strontium (Z=38)~\citep{Lodders2009}. The difference from the solar $r-$process residuals is shown in the bottom panel. The error bars show \textit{statistical} (random) errors.}
\label{fig:3}
\end{figure}

Figure~\ref{fig:3} and Table~\ref{tab:2} present the stars we used to compare to our nucleosynthesis calculations. These very metal poor stars ($\mathrm{[Fe/H] < -2}$) show high abundances of the light neutron-capture elements relative to heavy r-process elements between the second and third r-process peaks, as evidenced by their high ratio of Sr/Eu. The abundances in Figure~\ref{fig:3} are normalized to the solar\emph{r}-process strontium. There is an $\approx 0.5$~dex difference compared to the respective solar \textit{r}-process abundances (bottom panel Figure~\ref{fig:2}). The stars selected for this work are a subset of the star sample in~\cite{2022ApJ...935...27P}, ensuring that the stars have as many observed elements in the Z= 38-47 region as possible. For elements such as niobium and silver, there are not many published observations in the literature, compared to strontium-yttrium-zirconium, and that poses a challenge when comparing our nucleosynthesis theories to observations. Part of the reason for missing data is related to the wavelength of the strongest transitions of the heavy elements. Elements like silver and palladium show their strongest absorption lines in the near-UV/blue part of the spectrum ($<350$~nm) \citep{Hansen2012}, which is very hard to analyze due to line blending, and many spectrographs do not even cover this range due to low throughput of the signal. In Figure~\ref{fig:3}, technetium (Tc) is  missing. The reason is that Tc is radioactive with a lifteime that is many orders of magnitude shorter than the age of the weak r-process event that created the abundance pattern.  For comparison, while Sr shows much intrinsic stronger lines (for example at 407.7~nm), Y and Zr show features in the visual region around 500~nm which is covered by most large surveys.
The two stars with the most observed abundances in the Sr-Ag range in our sample are HD 122563 and HD 88609~\citep{2007ApJ...666.1189H}.

Observational uncertainties in abundances can be either statistical (random) or systematic. The first type originates mainly from line-to-line dispersion that can be caused by signal-to-noise ratio, uncertainties in line measurements or continuum placement, cosmic ray or sky line contamination, atomic data uncertainties, and other factors. Systematic errors are related to the atmospheric parameters used to determine the abundances ($\mathrm{T_{eff}}$, $\log g$, microturbulent velocities, metallicities) and the use of LTE or non-LTE (1D or multi-D)~\cite{2016MNRAS.463.1518A}. In Table~\ref{tab:2} we report only the \textit{random} uncertainties which we will use when fitting the abundance patterns in \S~\ref{sec:methods}.

\begin{deluxetable*}{lccccccccccl}\label{tab:2}
\tablecaption{Observations of metal--poor stars used in the present work in units of $\log \epsilon$. }
\tablewidth{\linewidth}
\tablehead{\colhead{Star} & \colhead{Sr} & \colhead{Y} & \colhead{Zr} & \colhead{Nb} & \colhead{Mo} & \colhead{Ru} & \colhead{Pd} &  \colhead{Ag} &  \colhead{Sr/Eu} & \colhead{Sr/Fe} &\colhead{[Fe/H]} }
\startdata
BD+42\_621 & 0.21(10) & -0.56(10) & 0.24(10) & \nodata & -0.40(10) & -0.53(10) &  \nodata &  -1.94(10) & 1.64 & -4.75 & -2.48\\
BD+06\_648 & 0.95(15) &0.02(15) &0.76(15) & \nodata &0.03(15) &-0.31(15) &-0.78(15) & \nodata & 2.33 & -4.38 & -2.11  \\
HD 23798 &0.86(15) &-0.04(15) &0.71(18) & \nodata &-0.11(15) &-0.17(15) &-0.74(15) & \nodata & 2.36 & -4.32 & -2.26 \\
HD 85773 & 0.00(16) &-0.96(18) &-0.23(16) & \nodata &-0.97(16) &-1.01(16) &-1.30(16)& \nodata & 2.08 & -4.82 & -2.62\\
HD 88609 & -0.20(12)& -0.98(10)& -0.24(16)&-1.72(12) &-1.00(12)& -0.91(12) & -1.35(12) & -2.03(12) & 2.69 & -4.62 & -2.87 \\
HD 107752 &-0.26(15) &-0.87(15) &-0.22(15) & \nodata &-0.90(15) &-0.96(15) &-1.35(15) & \nodata & 1.76 & -4.85 &-2.85 \\
HD 110184 &0.46(15) &-0.82(17) &-0.07(21) & \nodata &-0.70(15) &-0.85(15) &-1.22(15) & \nodata & 2.49 & -4.46 &-2.52 \\
HD 122563 & -0.12(14)&-0.93(9) &-0.28(16) &-1.48(14) &-0.87(14) &-0.86(14) & -1.31(14)&  -1.88(14) & 2.65 & -4.84 &  -2.72 \\
\enddata
\tablecomments{$\log \epsilon_X = \log \left( \frac{N_X}{N_H}\right) + 12$. The abundance data was taken by \cite{Hansen2012, 2014AA...568A..47H} (BD+42\_621), \cite{2017ApJ...837....8A} (BD+06\_648, HD 23798, HD 85773, HD 107752, HD 110184), \cite{2007ApJ...666.1189H, 2014AA...568A..47H} (HD 88609), \cite{2007ApJ...666.1189H} (HD 122563). The metallicities [Fe/H] are listed using the spectroscopic notation $\mathrm{[Fe/H] = \log \epsilon_{Fe,\star} - \log \epsilon_{Fe, \odot}}$. The uncertainties listed for Sr-Ag are \textit{statistical} (random).}
\end{deluxetable*}

\section{Methods}
\label{sec:methods}

We performed nucleosynthesis calculations using the \texttt{WinNet} reaction code~\citep{Winteler:2012, 2023arXiv230507048R}, and the same setup as in~\cite{2022ApJ...935...27P}, using the default reaction rates from the JINA Reaclib~\citep{Cyburt:2010} database and the $(\alpha, xn)$ reaction rates from~\cite{mohr2021astrophysical}, for the astrophysical conditions of Table~\ref{tab:1}.

In the following, we discuss our approach to match the astronomical observations using our extended library of astrophysical conditions (see \S~\ref{sec:astro}). Linear combinations of the different trajectories for the predicted abundance distribution were created using
\begin{equation}
    P = \sum_{i=1}^r w_i Y_i
\end{equation}
where the multiplication factors $w_i > 0$ represent the scaling applied to the abundance distribution $Y_i$, while $r$ is a free parameter indicating the number of distinct trajectories (conditions) selected for each combination from the sets presented in Table~\ref{tab:1}. The number of unique combinations of $N$ trajectories taken $r$ at a time is $C_r = N! / r! (N - r)!$, that is, for 2 and 3 selections out of the 50 trajectories in Table~\ref{tab:1}, there are 1,225 and 19,600 unique combinations, respectively. Each predicted abundance pattern $P$ is compared to the observed abundance pattern $O$ of the metal-poor stars in Table~\ref{tab:2}. Finding the weights that best fit the observational pattern for each combination is a linear regression (or least-squares) problem. To solve it, we need to:
\begin{equation}
   \mathrm{minimize}~ ||A w - O||^2
\end{equation}
where $A$ is a $k \times r$ matrix containing the $r$ individual abundance patterns $Y_i$ of the combination for $k$ elements, $w$ is the coefficient array (of length $r$) and $O$ is the observation pattern (of length $k$):

\begin{equation*}
A= \begin{bmatrix}
 Y_{11} & \cdots & Y_{1k}\\
\vdots & \ddots & \vdots \\
 Y_{r1} & \cdots & Y_{rk}
\end{bmatrix}  , w = \begin{bmatrix}
 w_1 \\
 \vdots \\
 w_N
\end{bmatrix}
\end{equation*}

\begin{figure*}[t!]
    \centering
    \includegraphics[width=.49\textwidth]{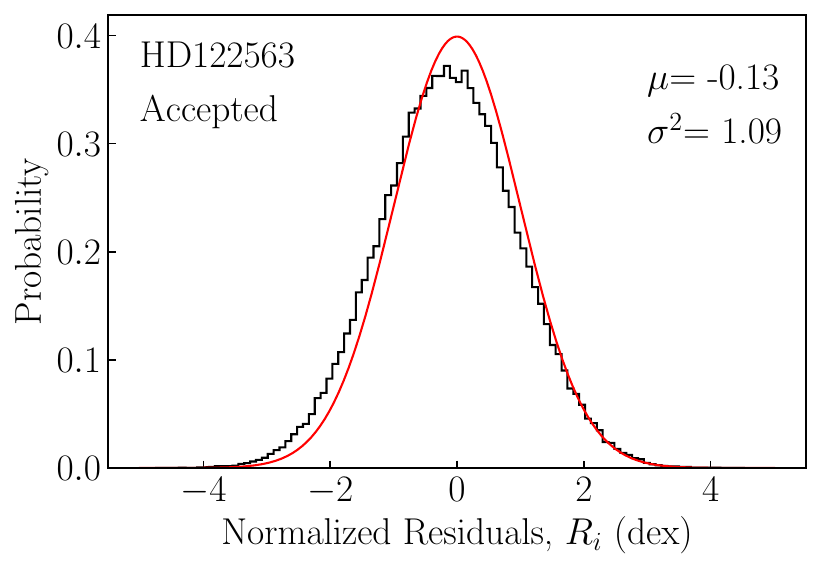}
    \includegraphics[width=.49\textwidth]{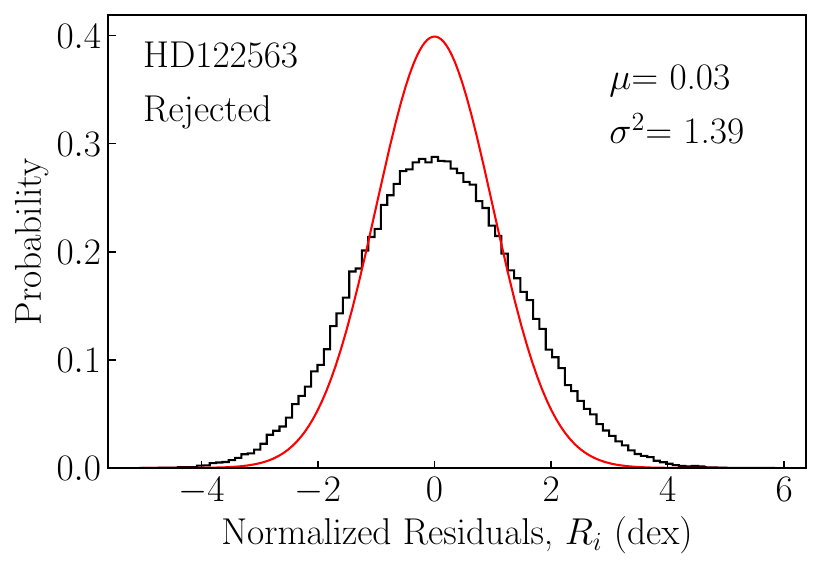}
    \caption{Distribution of normalized residuals for star HD 122563 using two trajectories ($r =2$). The red line shows the Normal distribution with $\mu = 0$ and $\sigma^2= 1$. The left panel shows an accepted fit, while the right shows a rejected one.}
\label{fig:residuals}
\end{figure*}

The coefficients $w_i$ self-normalize the predictions to the observations. We can define the relative contribution of each trajectory in the combination $F_i$ as follows:
\begin{equation}\label{eq:5}
    F_i = \frac{w_i}{\sum_{i=1}^{r} w_i}
\end{equation}
This way we can examine the most common thermodynamical properties ($Y_e$, entropy per baryon, and expansion timescale) that reproduce the elemental abundances in our star sample, and compare with hydrodynamical simulations (see \S~\ref{sec:results}).

To solve the linear regression problem, we used the \texttt{scikit-learn}~\citep{sklearn_api} package and weighted the sample according to the \textit{statistical (random)} observational uncertainty (Table~\ref{tab:2}).

The goodness of each fit to the observational data was assessed using a reduced $\chi^2_\nu$:
\begin{equation}\label{eq:chi2}
    \chi^2_\nu = \frac{1}{\nu} \sum_{i=1}^{k} \left( \frac{O(i)-P(i)}{\sigma(O(i))} \right)^2
\end{equation}
where $\sigma(O(i))$ is the observational \textit{statistical} uncertainty of element $i$, as shown in Table~\ref{tab:2} and $\nu = k-r$ are the degrees of freedom. In general, a lower $\chi^2_\nu$ value indicates a better model fit to the data, and values close to unity, assuming a reasonable number of data points, likely exhibit acceptable variance. Nevertheless, we are only fitting 6-8 points for each star, and each observational measurement has an inherent uncertainty $\sigma(O(i))$. To address these issues, we employed a Monte Carlo resampling technique by generating $10^4$ resampled datasets, according to the observational uncertainty of each star (Table~\ref{tab:2}). For each of these datasets, we recalculated the reduced $\chi^2_\nu$ value using Equation~\ref{eq:chi2}. This approach effectively simulates the impact of random errors and variability in the data. The result was a distribution of $\chi^2_\nu$ values, representing the range of fits we could expect due to data uncertainties. In the following, we shall report the median value of the resulted distributions, $\chi^2_{\nu,0.5}$.

Following the discussion in~\cite{2010arXiv1012.3754A}, the $\chi^2_\nu$ metric can only be used effectively for the true model having the true parameter values with \textit{a priori} known measurement errors. In that case, the normalized residuals:

\begin{equation}
    R_i =   \frac{O(i)-P(i)}{\sigma(O(i))}
\end{equation}
are normally distributed with mean $\mu = 0$ and variance $\sigma^2 = 1$ ($\sim \mathrm{Norm(0,1)}$). To accept a model according to that analysis, we performed an Anderson-Darling test (AD test)~\citep{Scholz1987} to the normalized residuals $R_i$ of each star and accepted those that followed $\sim \mathrm{Norm(0,1)}$ at the 5\% significance level. In Figure~\ref{fig:residuals} we show an example for an accepted and a rejected fit for star HD 122563 (for a full list of the accepted combinations, please see Tables~\ref{tab:n-combinations} and~\ref{tab:combinations-both}).

\section{Results} \label{sec:results}

In this section we present our results for the different combinations of astrophysical conditions: neutron-rich, proton-rich, and mixtures of both. For each star we shall report successful fits using the lowest number of trajectories $r$ necessary to obtain them.

\subsection{Neutron-rich conditions, $Y_e<0.5$}
\label{subsec:nrich}

\begin{figure}[ht!]
    \centering
    \includegraphics[width=.49\textwidth]{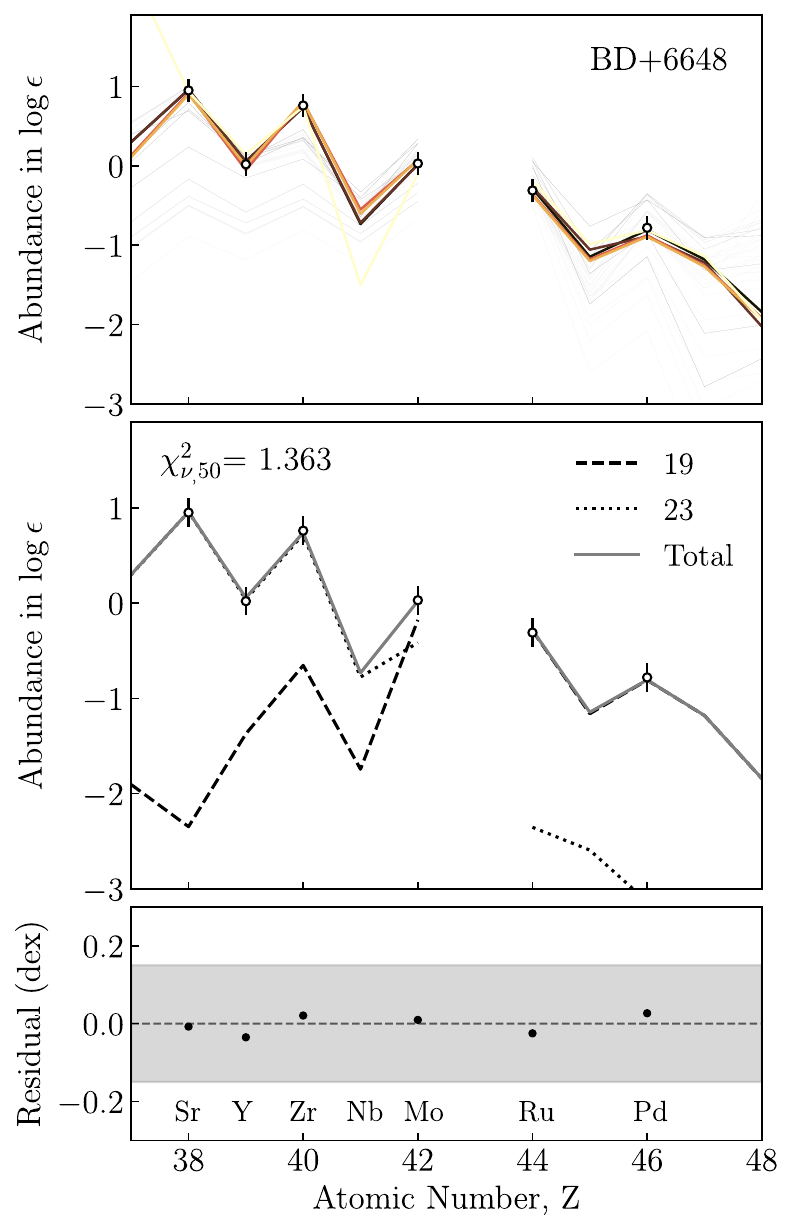}
    \caption{(Top) Fits for BD+06\_648 abundances using 2 neutron-rich conditions. Five colored lines represent accepted fits. Grey lines represent all the 630 unique 2 neutron-rich unique combinations from Table~\ref{tab:1} using the best weights $w_i$. (Middle) Decomposition of the best fit ($\chi^2_{\nu,50} = 1.363$) for BD+06\_648 using 2 neutron-rich conditions. (Bottom) Residuals of the best fit. The band shows the maximum random uncertainty of this star. See the text for details.}
\label{fig:5}
\end{figure}

The best combinations of two or three neutron-rich conditions (trajectories 1-36 from Table~\ref{tab:1}, $r=2,3$) that fit the elemental abundances are shown in Table~\ref{tab:n-combinations}. Note that according to the discussion we made in the above, we are looking for \textit{acceptable} fits. Figure~\ref{fig:5} - top shows a sample case of an accepted fit using 2 neutron-rich conditions ($r=2$) for metal-poor star BD+06\_648 using observational data from~\cite{2017ApJ...837....8A}. Only two components seem to reproduce the observations,  one for the $Z=38-42$ region and the other for the $Z=44-47$ abundances (Figure~\ref{fig:5} - middle). Such behavior is typical for our fits.

Of all the metal poor stars considered in Table~\ref{tab:2}, only stars HD 88609 and HD 122563 cannot provide acceptable fits for $r = 2, 3$. This discrepancy is likely attributed to their comparatively higher number of measured elemental abundances (8), which appears to adversely impact the fitting accuracy compared to other stars. For stars HD 23798 and BD+06\_648 we find acceptable fits for $r =2$. The rest of the stars, require $r=3$ fits to be considered acceptable (see Table~\ref{tab:n-combinations}). In general, neutron-rich trajectories are able to reproduce well the observational Sr-Zr abundances, but show some discrepancy in the Ru-Ag region. The most extreme example is HD 110184 that has the highest Sr/Ru and Sr/Mo ratios between  all the stars in our sample, and require at least 3 components to obtain an acceptable fit.

\begin{figure}[hbpt!]
    \centering
    \includegraphics[width=0.5\textwidth]{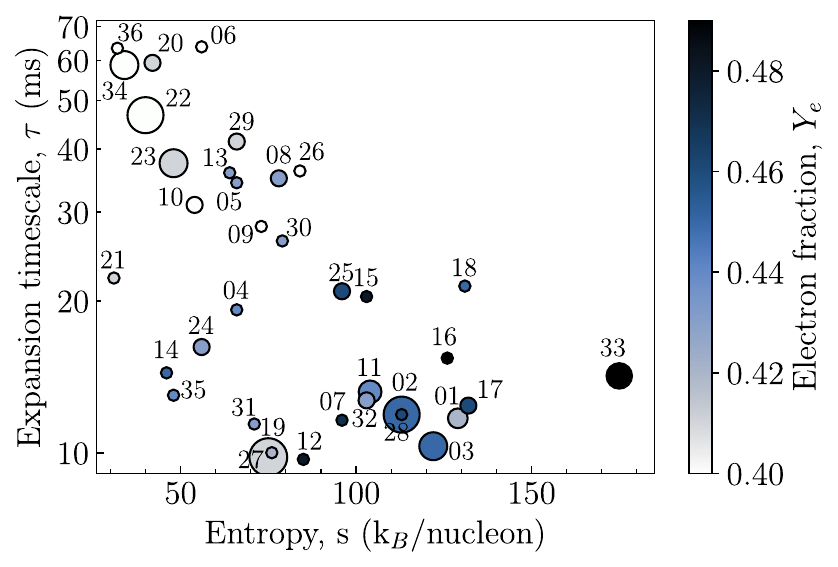}
    \caption{Same as Figure~\ref{fig:1} but showing only neutron-rich conditions. Sizes of the points are proportional to the number of appearances in accepted fits when combining only neutron-rich conditions.}
\label{fig:6}
\end{figure}

The neutron-rich trajectories that appear in the $r=2$ and $r=3$ accepted fits are shown in Figure~\ref{fig:6} in the relevant phase space. We find two clusters of conditions that are the most favorable in neutrino-driven ejecta: mainly $0.42 < Y_e <0.45$, high entropy ($s>80~k_B$/nucleon), and short expansion timescale ($ \tau < 15$~ms), and a smaller cluster with low $Y_e\approx0.40$, low entropy ($s<40~k_B$/nucleon), and high expansion timescale ($ \tau > 40$~ms).

\subsection{Proton-rich conditions, $Y_e>0.5$}
\label{sec:prich}

Similarly to the neutron-rich conditions, we performed calculations using combinations of two and three trajectories for the proton-rich conditions ($r = 2, 3$). The nucleosynthesis path of the $\nu p$-process is more robust compared to the weak \textit{r}-process~\citep{arcones2011production, arcones2014, nishimura2019uncertainties}. For this reason, we find much less variation in the abundance patterns of the different trajectories. The parameter that is most commonly used to assess the efficiency of the $\nu p$-process in producing heavy elements is the number ratio $\Delta_n$ Equation~\ref{eq:Delta_n}, as we discussed in \S~\ref{sec:astro}. \cite{nishimura2019uncertainties} have shown that a $\Delta_n \gtrsim 50$ leads to the production of $A>120$ nuclei, which we also confirm in our calculations.

\begin{figure}[hbpt!]
    \centering
    \includegraphics[width=.49\textwidth]{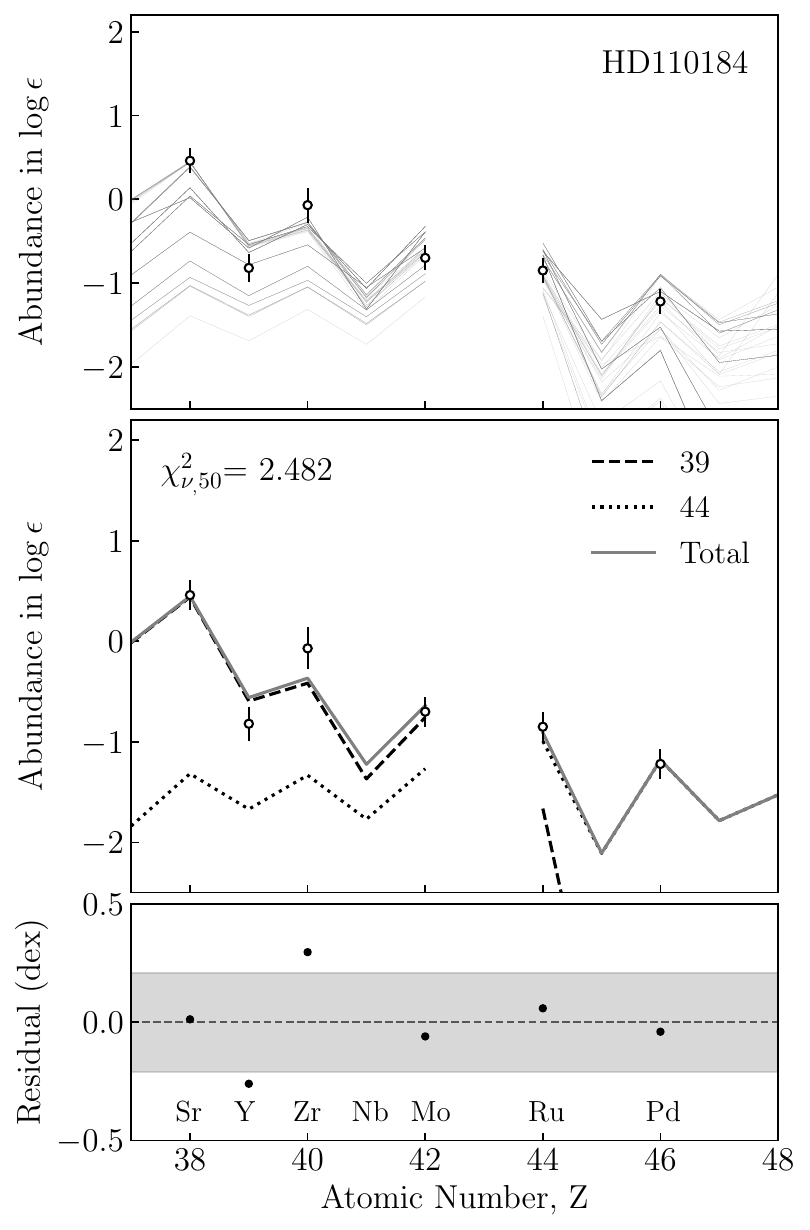}
    \caption{Same as Figure~\ref{fig:3} but using two proton-rich trajectories for star HD 110184. See the text for details.}
\label{fig:2p}
\end{figure}

None of the two proton-rich combinations produced an acceptable fit as shown by the large $\chi^2_{\nu,50}$ values in Table~\ref{tab:p-combinations}. The lowest $\chi^2_{\nu,50}$ (for star HD 110184) is 2.482 which is higher than any of the acceptable fits when using only neutron-rich conditions.  Nevertheless, in Figure~\ref{fig:2p}, we show the best combination for HD 110184. Trajectories 39 and 43 appear in all stars in Table~\ref{tab:p-combinations}, except in BD+42\_621. Interestingly, in the case of HD 107752, only one trajectory (43) produced the lowest $\chi^2_{\nu,50}$. We extended our analysis for $r = 3$ proton-rich trajectories to investigate whether we can obtain acceptable fits. Again, we did not obtain any acceptable fits, and for some stars, such as HD 23798 and HD 107752, we obtained the same results as in $r = 2$, showing that by increasing the $r$ value, the linear regression problem starts to degenerate.

The shortcoming of the proton-rich trajectories to produce acceptable fits can be explained by the variations in their Sr-Zr triplet abundances (see Figure~\ref{fig:2} compared to Figure~\ref{fig:3}). As we shall discuss in the following, the proton-rich trajectories can reproduce more accurately the Ru-Ag ($Z= 42-47$) region of the observed abundance pattern in VMP stars.

\subsection{Combination of neutron-rich and proton-rich conditions}
\label{sec:both}

Sophisticated multidimensional supernova simulations show that neutrino-driven ejecta are very complex and have both neutron- and proton-rich components~\citep[][for some recent examples]{2015ApJ...807L..31L, 2016ARNPS..66..341J, 2019MNRAS.482..351V, 2020MNRAS.491.2715B, 2021ApJ...921..113S, bollig2021self}.  For this reason, we decided to combine mixtures of both neutron-rich and proton-rich conditions.

Combining both neutron-rich and proton-rich conditions yields successful fits for all the stars in our sample for $r =2$, except for HD 110184 and HD 88609, for which we had to extend to $r = 3$ to find successful fits (see Table~\ref{tab:combinations-both} and the previous discussion). Out of the total 83 successful fits, 61 were mixtures of proton-rich and neutron-rich trajectories, while the rest were neutron-rich-only combinations. In Figure~\ref{fig:9} we present the results for all the stars in our sample, using two or three trajectories ($r = 2,3$) (only neutron-rich, only proton-rich, and a mixture of neutron-rich and proton-rich).

\begin{figure}[ht!]
    \centering
    \includegraphics[width=.49\textwidth]{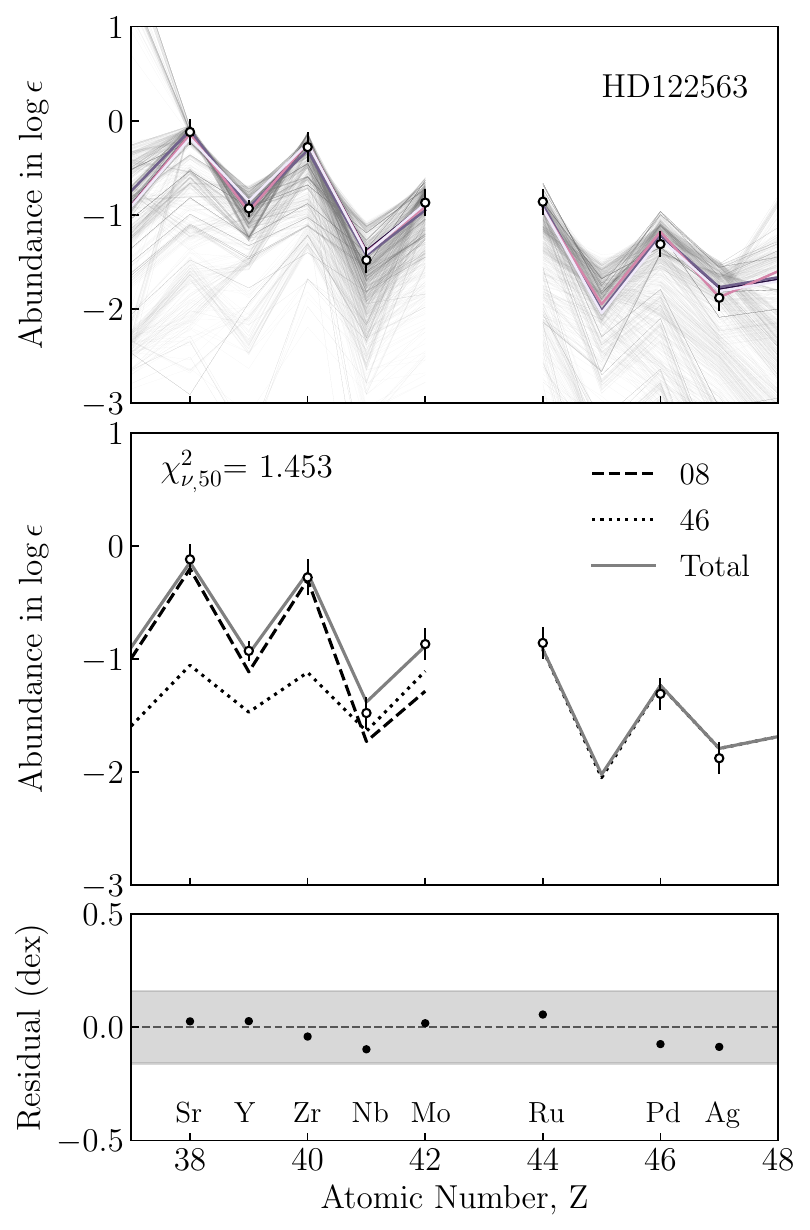}
    \caption{Same as Figure~\ref{fig:3} but with the combination one proton-rich and one neutron-rich condition for the star HD 122563. See the text for details.}
\label{fig-1n1p}
\end{figure}

\begin{figure*}[hbpt!]
    \centering
    \includegraphics[width=\textwidth]{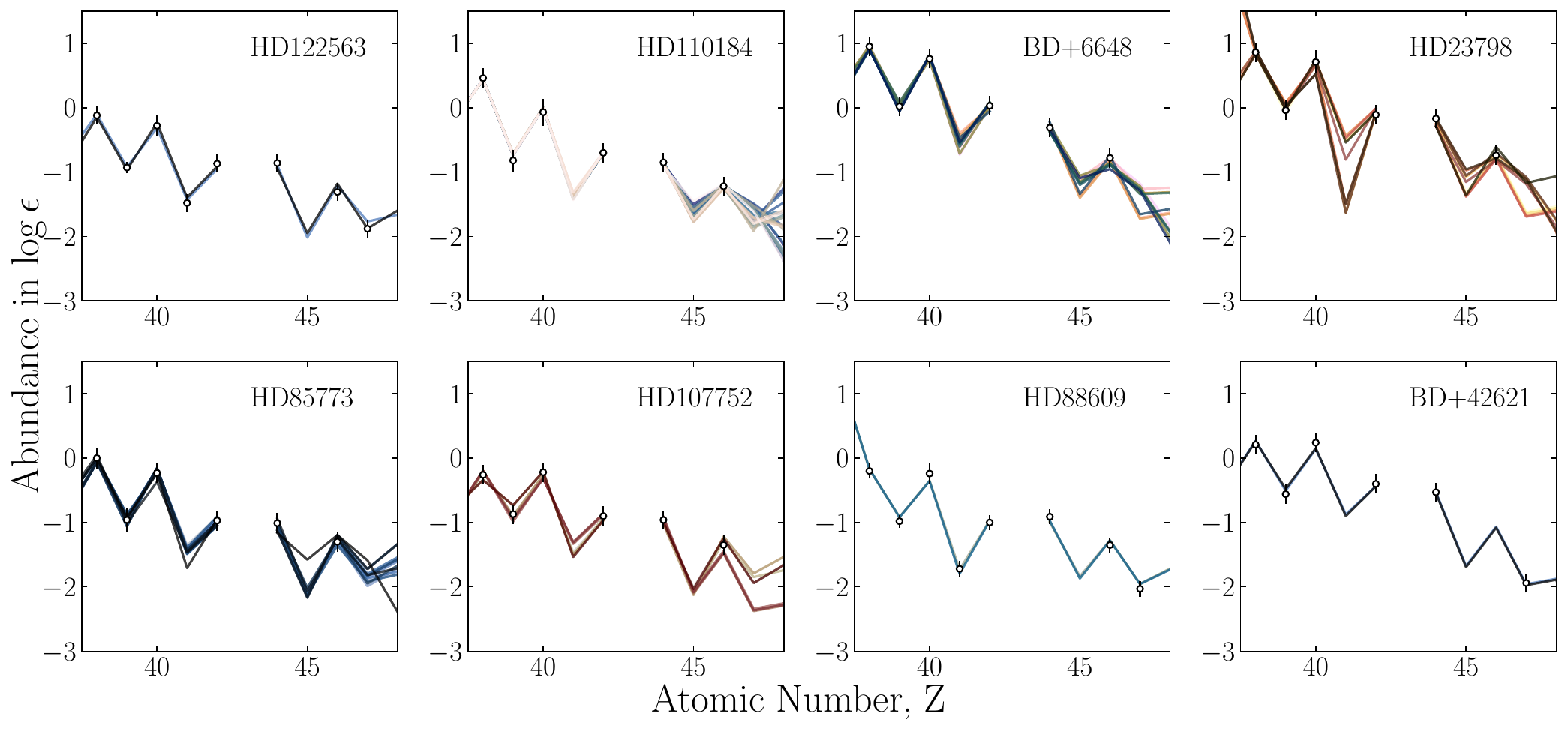}
    \caption{Abundance patterns of the accepted fits using $r= 2$ for the stars in our sample. Note that for HD 110184 and HD 88609 we use the $r= 3$ accepted fits. See the text for details.}
\label{fig:9}
\end{figure*}

\begin{figure*}[hptb!]
    \centering
    \includegraphics[width=\textwidth]{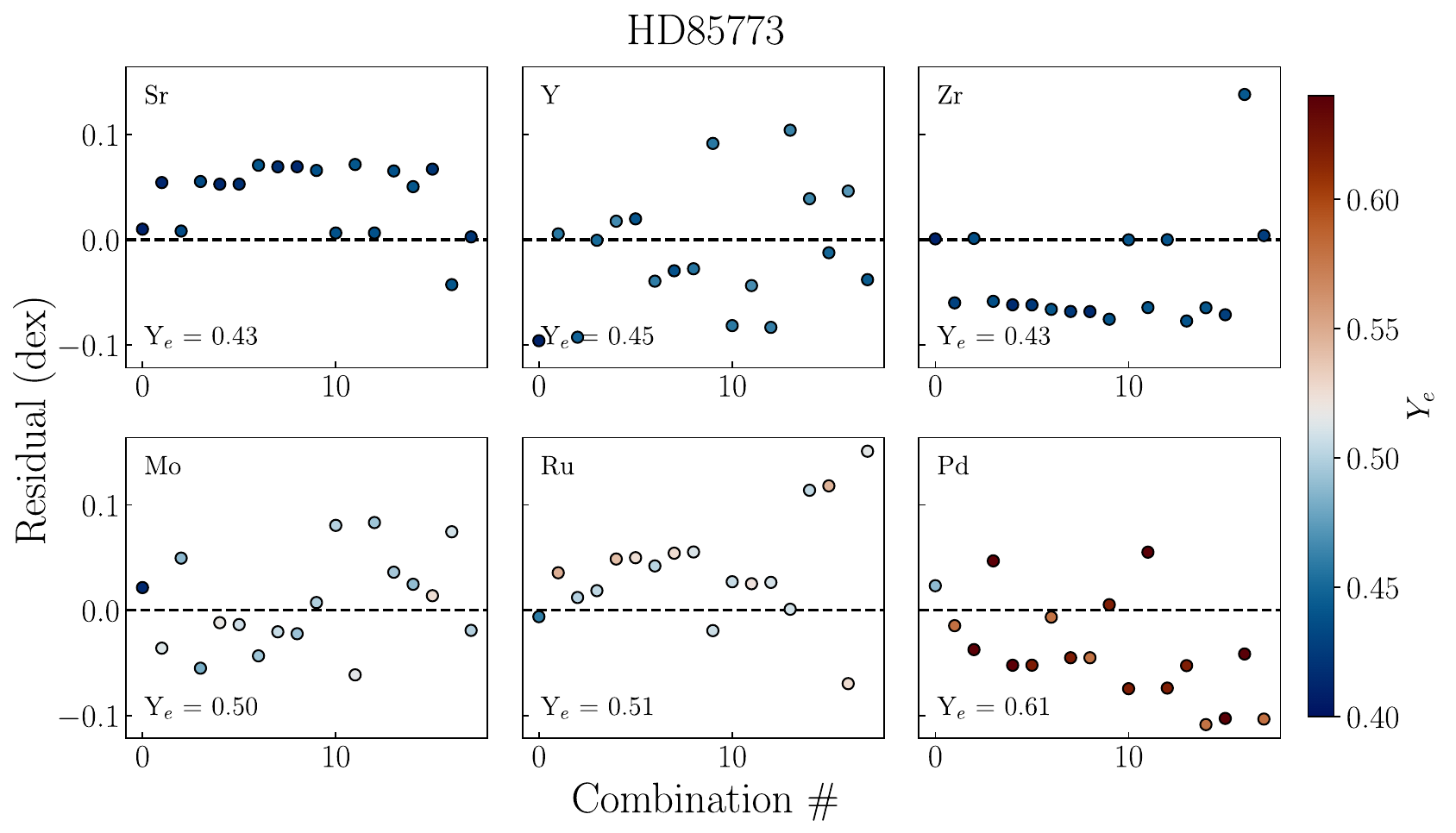}
    \caption{Residuals for the accepted fits (18) for HD 85773. Each panel shows the residuals for a particular element and the points are color-coded according to the weighted average $Y_e(Z) = (\sum_{i=1}^r w_i Y_i(Z) Y_{e,i})/(Y_{total}(Z))$ of the trajectories used for that element. It is evident that elements Sr, Y, Zr are better fitted using neutron-rich conditions. Mo needs a combination of neutron-rich and proton-rich material. Ru and Pd are best fitted using proton-rich conditions. See the text for details.}
\label{fig:individual}
\end{figure*}

In Figure~\ref{fig-1n1p}, we show an accepted fit for HD 122563. The proton-rich trajectories seem to reproduce more successfully the Ru-Ag ($Z= 44-47$) region. To further examine this, for each accepted fit we calculated the weighted average $Y_e(Z) = (\sum_{i=1}^r w_i Y_i(Z) Y_{e,i})/(Y_{total}(Z))$ for each element in $Z= 38-47$. In Figure~\ref{fig:individual} we show the residuals of the accepted fits for BD+06\_648 for each element color-coded them according to their respective $Y_e(Z)$. Sr-Zr is better produced by mean $\mathrm{Ye} = 0.43-0.45$. Mo requires a combinations of proton-rich and neutron-rich material, which yields a mean $Y_e = 0.50$. Finally, Ru and Pd need proton-rich conditions with a mean $Y_e = 0.51$ and $0.61$, respectively. This result is in agreement with~\cite{2017ApJ...837....8A} which argued that the decease observed in the abundances between Mo and Pd can be attributed to the weak \textit{r}-process. The behavior we described is found in all the successful fits which require one proton-rich and one neutron-rich trajectory and can also explain the reason why we were not able to find acceptable fits using only proton-rich trajectories.

\begin{figure}[ht!]
    \centering
    \includegraphics[width=0.5\textwidth]{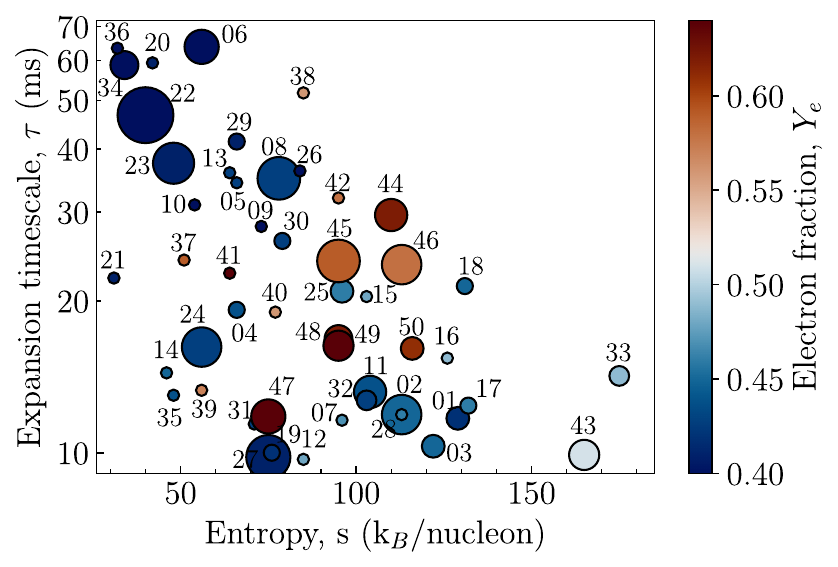}
    \caption{Same as Figure~\ref{fig:1}, the sizes of the points are proportional to the appearances in the accepted fits of Table~\ref{tab:combinations-both}.}
\label{fig:11}
\end{figure}

In Figure~\ref{fig:11}, we map these conditions in the relevant phase space to study the topology of the accepted fits under ccSNe conditions. There is a group of trajectories with moderate to high entropy ($40<s<113~k_B$/nucleon) and expansion timescale ($16<\tau<60$ ms) that participates in the majority of the accepted fits. A smaller group is located in low expansion timescale ($\tau < 12$~ms) and high entropy $60 < s < 120~k_B$/nucleon. An important conclusion is that the proton-rich trajectories dominate the accepted combinations, having much larger $R_i$ values (Equation~\ref{eq:5}), compared to the neutron-rich counterparts (see Table~\ref{tab:combinations-both} for a detailed list).

\section{Conclusions and Discussion}
\label{sec:conclusions}

In the context of the present work, we used thermodynamical trajectories of neutrino-driven ejecta from~\cite{bliss2018survey} (neutron-rich) and new proton-rich ones, spanning the relevant phase space ($\mathrm{Y_e}, s, \tau$), to fit the abundance patterns of the light neutron-capture elements between strontium and silver. For the first time, we combined both neutron-rich (weak \textit{r}-process) and proton-rich ($\nu p$-process) trajectories, sinnce a ccSNe is a complex phenomenon that produces a variety of conditions in its neutrino-driven phase, that can be described by these two processes.

We find that specific conditions are shown in multiple combinations in different stars (detailed list in Tables~\ref{tab:n-combinations} and~\ref{tab:combinations-both} and map in Figures~\ref{fig:6} and ~\ref{fig:11}), suggesting that they might be the dominant contributors for the production of $Z= 38-47$ elements in neutrino-driven outflows of ccSNe. Our results are consistent with recent multi-dimensional simulations, which show that the neutrino-driven ejecta are mainly proton-rich. However, the small neutron-rich component is crucial to reproduce the Sr-Zr abundances.

The analysis we have performed in the present work is based only on the abundances of \textit{elemental} and not \textit{isotopic} abundances. Since it is extremely difficult to discern different isotopes in the atmospheres of metal-poor stars, a viable alternative would be presolar stardust grains of ccSN origin, such as SiC~\citep{2013ApJ...767L..22P, 2018SciA....4.1054L}.

To enhance the robustness of our nucleosynthesis theories, further observations of metal-poor stars within the intriguing range from strontium to silver are essential. Additionally, comprehensive nucleosynthesis yields from multi-dimensional core-collapse supernova simulations will contribute significantly to advancing our understanding of their production in the Galaxy.

\begin{acknowledgements}
This work was supported by the Deutsche Forschungsgemeinschaft (DFG, German Research Foundation)—Project No. 279384907—SFB 1245,  the State of Hesse within the Research Cluster ELEMENTS (Project ID 500/10.006) and the National Science Foundation under Grant No. OISE-1927130 (IReNA). A.P is also supported by the U.S. Department of Energy, Office of Science, Office of Nuclear Physics, under Award Number DE-SC0017799 and Contract Nos. DE-FG02-97ER41033 and DE-FG02-97ER41042. The network calculations were performed on the GSI Virgo HPC cluster. FM and HS are supported by the US National Science Foundation under award numbers PHY-1913554 and PHY 14-30152 (JINA-CEE) and NSF Grant No. 2209429. CJH acknowledges the European Union’s Horizon 2020 research and innovation program under grant agreement No. 101008324 (ChETEC-INFRA).
\end{acknowledgements}

    \software{ \texttt{h5py}~\citep{collette_python_hdf5_2014}, \texttt{IPython}~\citep{PER-GRA:2007}, \texttt{Jupyter}~\citep{Kluyver2016jupyter},\texttt{matplotlib}~\citep{Hunter:2007}, \texttt{numpy}~\citep{harris2020array}, \texttt{pandas}~\citep{reback2020pandas}, Scientific colour maps~\citep{crameri_fabio_2021_5501399}, \texttt{sklearn}~\citep{sklearn_api}}


\bibliography{bibliography}{}
\bibliographystyle{aasjournal}



\appendix

\startlongtable
\begin{deluxetable}{cccc}\label{tab:n-combinations}
\tablecaption{Best combinations of two or three neutron-rich conditions that fit the elemental abundances of our star sample. The relative contributions $F_i$ and the $\chi^2_{\nu,50}$ are also shown. The combinations below the line are the best but not acceptable fits. See the text for details.}
\tablehead{\colhead{Star} & \colhead{Trajectories} & \colhead{Relative contribution, $F_i$} &  \colhead{$\chi^2_{\nu,50}$} }
\startdata
BD+06\_648 & 19, 23 & 0.1165, 0.8835 & 1.363 \\
BD+06\_648 & 02, 23 & 0.1999, 0.8001 & 1.497 \\
BD+06\_648 & 19, 24 & 0.0908, 0.9092 & 1.564 \\
BD+06\_648 & 08, 19 & 0.9184, 0.0816 & 1.595 \\
BD+06\_648 & 02, 34 & 0.0036, 0.9964 & 1.848 \\
HD 23798 & 19, 34 & 0.0023, 0.9977 & 1.485 \\
HD 23798 & 02, 34 & 0.0044, 0.9956 & 1.518 \\
HD 23798 & 19, 23 & 0.1358, 0.8642 & 1.676 \\
HD 23798 & 02, 23 & 0.2317, 0.7683 & 1.808 \\
BD+42\_621 & 11, 29, 34 & 0.5793, 0.4207, 0.9865& 1.203 \\
HD 85773 & 02, 03, 23 & 0.4785, 0.5215, 0.6919& 1.154 \\
HD 85773 & 10, 20, 33 & 0.1287, 0.8713, 0.4845& 1.185 \\
HD 85773 & 02, 33, 34 & 0.1345, 0.8655, 0.9806& 1.189 \\
HD107752 & 02, 03, 25 & 0.5914, 0.4086, 0.7829& 1.272 \\
HD 110184 & 02, 03, 22 & 0.1684, 0.0672, 0.7644 & 1.918 \\
HD 110184 & 01, 19, 22 & 0.0432, 0.1126, 0.8442 & 1.930 \\
HD 110184 & 02, 17, 22 & 0.1354, 0.1162, 0.7484 & 1.932 \\
HD 110184 & 19, 22, 33 & 0.0542, 0.5738, 0.3720 & 1.946 \\
HD 110184 & 03, 11, 22 & 0.0946, 0.1223, 0.7831 & 1.978 \\
HD 110184 & 19, 22, 32 & 0.0908, 0.8587, 0.0505 & 1.951 \\
HD 110184 & 03, 19, 22 & 0.0636, 0.0991, 0.8373 & 1.959 \\
HD 110184 & 11, 22, 33 & 0.0584, 0.5010, 0.4406 & 1.964 \\
\hline
HD88609 & 19, 23 & 0.2209, 0.7791 & 2.678 \\
HD88609 & 10, 19, 34 & 0.2902, 0.7098, 0.9951& 2.684 \\
HD122563 & 19, 23 & 0.2409, 0.7591 & 2.289  \\
HD122563 & 10, 19, 36 & 0.3469, 0.6531, 0.9978& 1.982 \\
\enddata
\end{deluxetable}

\startlongtable
\begin{deluxetable}{cccc}\label{tab:p-combinations}
\tablecaption{Best combinations of two  and three proton-rich conditions that fit the elemental abundances of our star sample. The relative contributions $F_i$ and the reduced $\chi^2_{\nu,50}$ are also shown. None of these combinations is an acceptable fit.}
\tablehead{\colhead{Star} & \colhead{Trajectories} & \colhead{Relative contribution, $F_i$} &  \colhead{$\chi^2_{\nu,50}$} }
\startdata
HD 122563 & 39, 43 & 0.4989, 0.5011 & 3.157 \\
HD 122563 & 39, 43, 46 & 0.5991, 0.3969, 0.0040& 2.344 \\
HD 110184 & 39, 44 & 0.9981, 0.0019 & 2.482 \\
HD 110184 & 39, 44, 46 & 0.9977, 0.0013, 0.0010& 3.232 \\
BD+06\_648 & 39, 43 & 0.8080, 0.1920 & 3.340 \\
BD+06\_648 & 39, 43, 49 & 0.8080, 0.1920, 0.0000& 4.427 \\
HD 23798 & 39, 43 & 0.7750, 0.2250 & 3.002 \\
HD 85773 & 40, 49 & 0.9869, 0.0131 & 3.236 \\
HD 85773 & 39, 43, 49 & 0.7947, 0.2037, 0.0015& 3.826 \\
HD 107752 &  43 &  1.0000 & 2.730 \\
HD 88609 & 39, 43 & 0.5111, 0.4889 & 3.155 \\
HD 88609 & 39, 43, 50 & 0.5696, 0.4275, 0.0029& 3.446 \\
BD+42\_621 & 41, 49 & 0.9876, 0.0124 & 7.254 \\
BD+42\_621 & 40, 41, 46 & 0.3999, 0.5731, 0.0271& 9.332
\enddata
\end{deluxetable}

\startlongtable
\begin{deluxetable}{cccc}\label{tab:combinations-both}
\tablecaption{Best combinations of two or three neutron-rich and proton-rich conditions that fit the elemental abundances of our star sample. The relative contributions $F_i$ and the $\chi^2_{\nu,50}$ are also shown. All of them are accepted fits.}
\tablehead{\colhead{Star} & \colhead{Trajectories} & \colhead{Relative contribution, $F_i$} &  \colhead{$\chi^2_{\nu,50}$} }
\startdata
HD 122563 & 08, 46 & 0.0302, 0.9698 & 1.453 \\
HD 122563 & 23, 46 & 0.0231, 0.9769 & 1.490 \\
BD+06\_648 & 06, 43 & 0.0015, 0.9985 & 1.374 \\
BD+06\_648 & 24, 45 & 0.0478, 0.9522 & 1.395 \\
BD+06\_648 & 08, 45 & 0.0536, 0.9464 & 1.436 \\
BD+06\_648 & 24, 43 & 0.0024, 0.9976 & 1.484 \\
BD+06\_648 & 19, 24 & 0.0908, 0.9092 & 1.569 \\
BD+06\_648 & 08, 43 & 0.0027, 0.9973 & 1.563 \\
BD+06\_648 & 06, 45 & 0.0313, 0.9687 & 1.581 \\
BD+06\_648 & 23, 45 & 0.0407, 0.9593 & 1.583 \\
HD 23798 & 06, 45 & 0.0238, 0.9762 & 1.409 \\
HD 23798 & 23, 45 & 0.0294, 0.9706 & 1.496 \\
HD 23798 & 24, 45 & 0.0362, 0.9638 & 1.492 \\
HD 23798 & 06, 43 & 0.0013, 0.9987 & 1.886 \\
HD 23798 & 08, 47 & 0.0796, 0.9204 & 1.901 \\
HD 23798 & 08, 44 & 0.1112, 0.8888 & 2.111 \\
HD 85773 & 23, 46 & 0.0384, 0.9616 & 1.418 \\
HD 85773 & 24, 47 & 0.0475, 0.9525 & 1.415 \\
HD 85773 & 23, 47 & 0.0413, 0.9587 & 1.429 \\
HD 85773 & 24, 46 & 0.0443, 0.9557 & 1.443 \\
HD 85773 & 24, 44 & 0.0648, 0.9352 & 1.445 \\
HD 85773 & 24, 48 & 0.0564, 0.9436 & 1.447 \\
HD 85773 & 08, 48 & 0.0619, 0.9381 & 1.460 \\
HD 85773 & 08, 44 & 0.0710, 0.9290 & 1.478 \\
HD 85773 & 08, 47 & 0.0522, 0.9478 & 1.453 \\
HD 85773 & 06, 46 & 0.0305, 0.9695 & 1.489 \\
HD 85773 & 23, 48 & 0.0492, 0.9508 & 1.496 \\
HD 85773 & 08, 46 & 0.0487, 0.9513 & 1.504 \\
HD 85773 & 23, 44 & 0.0567, 0.9433 & 1.503 \\
HD 85773 & 06, 47 & 0.0330, 0.9670 & 1.529 \\
HD 85773 & 24, 49 & 0.0627, 0.9373 & 1.613 \\
HD 85773 & 08, 49 & 0.0686, 0.9314 & 1.642 \\
HD 85773 & 25, 47 & 0.0914, 0.9086 & 1.731 \\
HD 85773 & 23, 49 & 0.0549, 0.9451 & 1.775 \\
HD 107752 & 25, 46 & 0.0558, 0.9442 & 1.436 \\
HD 107752 & 08, 46 & 0.0293, 0.9707 & 1.688 \\
HD 107752 & 08, 45 & 0.0161, 0.9839 & 1.700 \\
HD 107752 & 06, 45 & 0.0099, 0.9901 & 1.702 \\
HD 107752 & 24, 45 & 0.0140, 0.9860 & 1.715 \\
HD 107752 & 29, 43 & 0.0004, 0.9996 & 1.870 \\
HD 107752 & 04, 45 & 0.0253, 0.9747 & 1.887 \\
HD 107752 & 30, 43 & 0.0006, 0.9994 & 1.957 \\
BD+42\_621 & 06, 45 & 0.0115, 0.9885 & 1.660 \\
BD+42\_621 & 24, 45 & 0.0165, 0.9835 & 1.757 \\
HD 110184 & 11, 22, 49 & 0.0228, 0.1566, 0.8206 & 1.895 \\
HD 110184 & 02, 22, 50 & 0.0342, 0.1556, 0.8102 & 1.902 \\
HD 110184 & 02, 22, 32 & 0.1546, 0.7930, 0.0524 & 1.907 \\
HD 110184 & 19, 22, 50 & 0.0210, 0.1781, 0.8008 & 1.911 \\
HD 110184 & 01, 02, 22 & 0.0490, 0.1912, 0.7598 & 1.917 \\
HD 110184 & 19, 22, 49 & 0.0267, 0.2327, 0.7406 & 1.920 \\
HD 110184 & 11, 22, 44 & 0.0211, 0.1572, 0.8218 & 1.936 \\
HD 110184 & 02, 22, 49 & 0.0442, 0.2079, 0.7479 & 1.937 \\
HD 110184 & 19, 22, 48 & 0.0232, 0.2106, 0.7662 & 1.939 \\
HD 110184 & 11, 22, 50 & 0.0182, 0.1179, 0.8639 & 1.942 \\
HD 110184 & 22, 27, 44 & 0.1328, 0.0108, 0.8564 & 1.950 \\
HD 110184 & 02, 22, 48 & 0.0387, 0.1899, 0.7714 & 1.958 \\
HD 110184 & 11, 22, 47 & 0.0148, 0.1158, 0.8695 & 1.973 \\
HD 110184 & 19, 22, 44 & 0.0264, 0.2380, 0.7356 & 1.981 \\
HD 110184 & 19, 22, 47 & 0.0202, 0.1846, 0.7952 & 1.988 \\
HD 88609 & 11, 34, 46 & 0.0016, 0.7382, 0.2602& 1.922 \\
HD 88609 & 18, 34, 46 & 0.0016, 0.7127, 0.2858 & 2.038
\enddata
\end{deluxetable}

\end{document}